\documentclass[conference]{IEEEtran}

\usepackage{cite}
\usepackage[dvips]{graphicx}
\usepackage{multirow}

\begin{document}

\title{Economical and Efficient Big Data Sharing with i-Cloud}

\author{\IEEEauthorblockN{Thepparit Banditwattanawong}
\IEEEauthorblockN{Masawee Masdisornchote}
\IEEEauthorblockA{School of Information Technology, Sripatum University\\Bangkok, Thailand}
\and
\IEEEauthorblockN{Putchong Uthayopas}
\IEEEauthorblockA{Computer Engineering Department\\Kasetsart University\\Bangkok, Thailand}}

\maketitle

\begin{abstract}
Big data can be hosted on cloud and being shared distributedly through cloud services in an unprecedented volume, variety and velocity. This causes not only cloud network congestions and delayed cloud services but also increases in public cloud data-out charges. Client-side cloud cache alleviates these problems. Furthermore, cloud cache must be aware of nonuniform data-out costs when big data is stored in hybrid clouds built with different public cloud providers. Deploying i-Cloud approach as the core mechanism of cloud cache could save data-out cost up to 14.78\% or 4,425 USD saved per annum based on our representative scenario, and delivered 17.24\% byte-hit, 17.96\% delay-saving and 29.33\% cache hit outperforming LRU, GDSF and LFU-DA approaches. A main finding is that i-Cloud, learning uniform cost patterns, could perform well against nonuniform cost environment.
\end{abstract}


\section{Introduction}
\label{Introduction}
Big data such as social media contents, archive of high-definition videos gathered via ubiquitous information-sensing devices and scientific data could be acquired and stored within an organization's external cloud(s) and distributedly retrieved by staffs or customers via cloud services offered by the organization. This leads to the downstream bandwidth saturation of network connection between external cloud and big data consumer premise, long-delayed cloud service responsiveness and importantly increases in external cloud data-out charge imposed by public cloud provider \cite{AWS,GAE,WA}. The significance of the last problem could be realized through the following representative scenario (which is also referred to throughout this paper): an enterprise utilizing big data residing in clouds by transferring it through 10 Gbps Metro Ethernet with 25\% average downstream bandwidth utilization for 8 work hours a day, and 260 workdays per year requires the total amount of cloud data-out transfer 190.43 TB per month. This data transfer volume can be translated as 29,933 USD per month based on the weighted average cost 0.1535 USD per GB of Google Cloud Storage's network egress charge in Asia-Pacific region as of September 2013 \cite{GAE}.

\begin{table*}[!tp]
\renewcommand{\arraystretch}{1.25}
\caption{Characteristics of preprocessed traces.}
\label{traces}
\centering
\begin{footnotesize}
\begin{tabular}{c c c c c}
\hline
\multirow{2}{*}{Feature} & \multicolumn{2}{c}{BO} & \multicolumn{2}{c}{NY}\\ \cline{2-5}
\multirow{2}{*} & 15 days & 31 days & 15 days & 31 days\\
\hline
Total requests & 352,224 & 639,187 & 639,199 & 1,311,880\\
\hline
Total requested bytes & 2,294,688,191 & 4,149,211,314 & 6,499,655,874 & 17,067,821,671\\
\hline
Total unique objects & 181,624 & 323,979 & 290,851 & 593,365\\
\hline
Maximum bytes of & 1,386,970,321 & 2,262,144,480 & 4,791,008,825 & 10,801,010,237\\
total unique objects & & & &\\
\hline
\end{tabular}
\end{footnotesize}
\end{table*}

\begin{figure}[!t]
\centering
\includegraphics[width=3.3in]{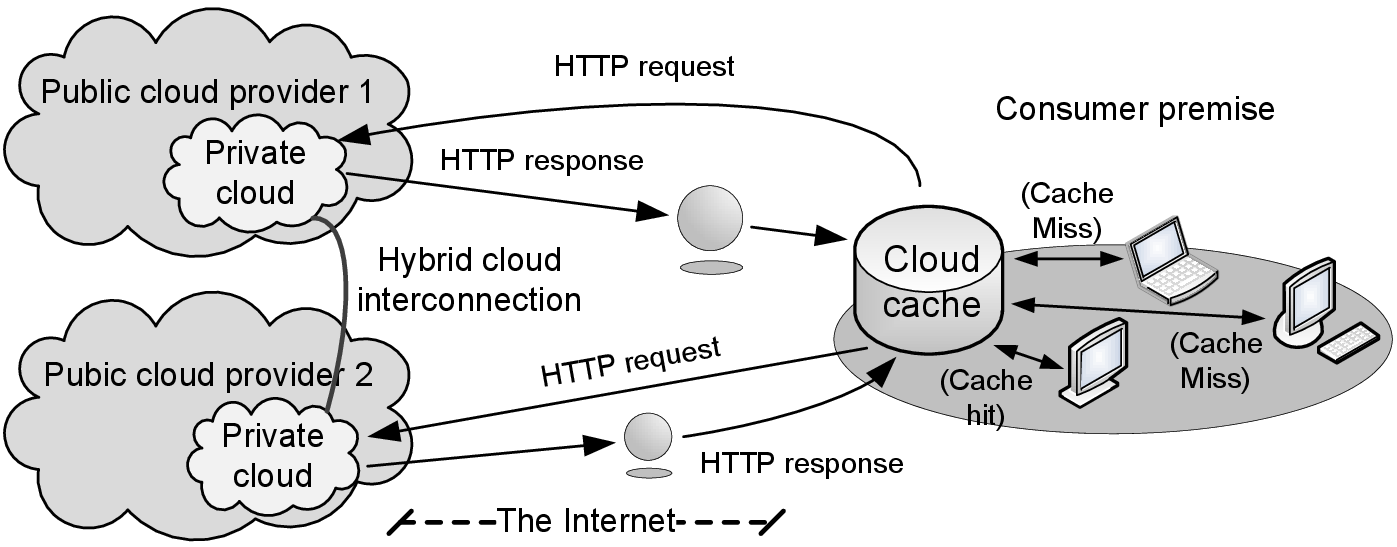}
\caption{Cloud cache deployment in a hybrid cloud scenario.}
\label{Cloudcachedeploymentscenario}
\end{figure}

The sharings of big data can be conducted in an economical and network-friendly manner by using client-side cloud cache. Client-side cloud caches are located in or nearby user premise in the form of enterprise-level shared cache, personal web-browser cache or local user-application cache. Fig.\,\ref{Cloudcachedeploymentscenario} demonstrates the deployment scenario of a shared cloud cache where HTTP requests to external hybrid cloud are proxied by a cloud cache, which in turn replies with the valid copies of the requested big data objects either from its local cache repository (i.e., cache hits) or by retrieving updated copies from the cloud (i.e, cache misses). Cloud caches inherit the capabilities of traditional forward web caching proxies since cloud data is also delivered by using the same set of HTTP/TCP/IP protocol stacks as in WWW. Unavoidably, the same problem as in web caching proxies also exists in cloud caches that is caching entire remote data in local cache is not economically plausible, thus cache eviction approach is mandatory for cloud caches. When the big-data hosting cloud is a kind of hybrid, which employs different public cloud providers for risk management purpose, different data-out charge rates potentially apply to data-out costs and must be aware of by cache eviction approach for economical performance optimization.

\section{Related works}
There are numerous cache eviction approaches in present existence such as \cite{Podlipnig, nnpcr1, nnpcr, Ali, SWORM, AWC}. They have been extensively investigated in our previous works \cite{cloudcache1}. To recap, none of them aims for big data and cloud computing for two main reasons. First, those approaches evict big objects to optimize hit rates rather than byte-hit and delay-saving ratios, crucial to the scalability of cloud-transport infrastructures and the responsiveness of cloud computing services, respectively. Second, they do not support multiple public-cloud data-out charges, thus neither improve cloud consumer-side economy nor support hybrid cloud deployment. i-Cloud approach, originally proposed in \cite{cloudcache4}, extends its prior nonintelligent versions \cite{cloudcache1, cloudcache2, cloudcache3} by integrating an artificial neural network (ANN) to automate an algorithmic parameter self-tuning for workload adaptability. Its performances have been studied without comparing with the other well-known approaches and based on the totally uniform-cost circumstances of both ANN training and deployment phases. 

This paper presents the new and comparative performance behaviors of i-Cloud and three well-known approaches by emulating a hybrid cloud as a testing environment where economical costs offered by two public cloud providers are nonuniform. The main objective of doing this is to observe the performances of i-Cloud that has learned uniform cost patterns but is deployed against a nonuniform cost environment. A minor objective is to show how much i-Cloud outperforms the other approaches when data-out charge rates are nonuniform. The findings of these observations would convince users of i-Cloud performances when deploying cloud cache for a single private cloud at the beginning that later evolves to a hybrid cloud according to new business requirements.

\section{Original data sets}
\label{Original trace data sets}
To understand the following sections, it is necessary to clarify our original trace data sets from which the other types of data sets used in our study were derived. The trace data sets were provided by IRCache project \cite{ircache} in the form of raw HTTP traces representing the object request streams of two separate user-community behaviors: a 31-day BO trace was gathered from a user community in Boulder from 16th August to 15th September 2012, a 31-day NY trace was collected from the other user community in New York from 16th July to 15th August 2012.

To emulate cloud computing request streams of the same consumer organization, these raw traces must be adjusted with realistic assumptions. For this reason, each of these traces was preprocessed to extract only the object references of 50 most popular domain names (to emulate the total number of HTTP domains administrated within the first author's university). Omitting this preprocessing step was translated that a single consumer organization owns an impractically large number of domain names hosted on its own cloud(s). All references representing dynamic object requests were not excluded during preprocessing stage to reflect actual caching performances against all types of requests. (That is why our performance results were not so high as was reported in some of the related works.) We found that the extracted references reflected cloud traffics or big data as containing both cloud service requests to, for instance, Facebook, Youtube, Twitter SaaSes, and Mediafire IaaS, and WWW requests, which were assumed to go to cloud-hosted web servers. To make the preprocessed traces self-contained, additional preprocessings were the removals of unused fields and appending object expiration times as a new field to all object references. The expiration times were figured out based on the following heuristic rules with an assumption that object creators set object expiration times deliberately:

\begin{itemize}
\item Rule 1: by scanning down all references within a trace in timestamp order, object expired immediately after its size was found changed from its last reference.
\item Rule 2: as long as the size of object remained unchanged, its lifetime was extended to its last request as appeared in a trace.
\item Rule 3: object apparent only once throughout a trace expired suddenly after its use. (This object tends to be a kind of dynamic one, which is by default not cached by Squid. Thus, this rule is prescribed so.) 
\end{itemize}

Once the one-month BO and NY traces had been preprocessed, the references belonging to the first 15 days of each trace were duplicated into a new trace resulting in a 15-day BO trace and a 15-day NY trace. This allowed objects referenced in both 15-day traces to have maximum one-month lifespans (as a result of applying the heuristic rule to the one-month traces) rather than merely 15 days, and avoided false positives for static objects made by the heuristic rule 3. The final preprocessing results are characterized in Table\,\ref{traces}. Notice that the last feature must be denoted as the maximum bytes (of total unique objects) since the sizes of several unique objects enlarged over a reference stream as detected in each trace.

\section{Economical and Technical Performance Measurement}
\label{Performance metrics}
The algorithm design and results of i-Cloud have been described in standard performance metrics \cite{Podlipnig}, defined as follows. For an object $i$, $byte{-}hit~ratio = \sum_{i=1}^{n}s_{i} h_{i}/\sum_{i=1}^{n}s_{i} r_{i}$, $delay{-}saving~ratio = \sum_{i=1}^{n}l_{i} h_{i}/\sum_{i=1}^{n}l_{i} r_{i}$ and $hit~rate = \sum_{i=1}^{n}h_{i}/\sum_{i=1}^{n}r_{i}$ where \textit{s$_{i}$} is the size of \textit{i}, \textit{h$_{i}$} is how many times a valid copy of \textit{i} is fetched from cache, \textit{r$_{i}$} is the total number of requests to \textit{i}, and \textit{l$_{i}$} is the loading latency of \textit{i} from cloud. In addition, cost-saving ratio \cite{cloudcache1} was also used to capture the economical performances of our studied approaches. The metric measures how much money can be saved by serving the valid copies of requested objects from cache. It is expressed as below. Given an object \textit{i},

\begin{equation} \label{eq4}
cost{-}saving~ratio = \frac{\sum_{i=1}^{n}c_{i} s_{i} h_{i}}{\sum_{i=1}^{n}c_{i} s_{i} r_{i}}
\end{equation}

\setlength{\parindent}{0in}where \textit{c$_{i}$} is the data-out charge rate or monetary cost for loading \textit{i} from cloud. This metric is particularly useful for a hybrid cloud or a nonuniform-cost model (Cf. Section \ref{Monetary cost models}).\parindent 1.0em

\section{$i$-Cloud Approach}
\label{i-Cloud Approach}
The design goals of i-Cloud are cloud economy, scalability and responsiveness that can be realized by optimizing cost-saving, byte-hit and delay-saving ratios, respectively. Hit rate has gained declining impact  since these days with globally available broadband network infrastructures, it is perceived that loading remote small objects is fast as if they were fetched from user locus whereas the problems still insist on retrieving big objects over the network (more details are available in \cite{cloudcache1,cloudcache2,cloudcache3}). 

Table\,\ref{i-CloudAlgorithm} presents the pseudo code of i-Cloud algorithm in details. The main principle behind the scene of i-Cloud is contemporaneous proximity \cite{ieej}. When cache eviction is needed, i-Cloud is invoked. It first formulates a cluster of in-cache least-recently-used (lru) objects as many as instructed either by a $window~size$ parameter or a required cache space (which is at least a requested missing object's size and controlled by Squid's low and high watermarks \cite{Squid}) depending on which one is larger, it will be chosen. Once the cluster of lru objects has been formed, i-Cloud quantifies a profit associated with each object inside the cluster as follows: given an object \textit{i}, $profit_{i} = s_{i} \cdot c_{i} \cdot l_{i} \cdot f_{i} \cdot TTL_{i}$ where \textit{s$_{i}$} is the size of \textit{i}, \textit{c$_{i}$} is data-out charge rate for loading \textit{i}, \textit{l$_{i}$} is latency for loading \textit{i}, \textit{f$_{i}$} is the access frequency of \textit{i}, and \textit{TTL$_{i}$} is the remaining lifespan of \textit{i}. (This profit formula is justified in \cite{cloudcache1}.) An object with least profit is evicted first from cache. This object eviction process is repeated on the next least profitable objects in the cluster until gaining enough cache room.

\begin{table}[!t]
\renewcommand{\arraystretch}{1.2}
\caption{i-Cloud algorithm}
\label{i-CloudAlgorithm}
\begin{footnotesize}
\begin{tabular}{l}
\hline
\textbf{algorithm:} \texttt{i-Cloud}\\
\hline
\textbf{input variables:}\\
$~~~~~~~cd$~~~~/*cache database (recency-keyed min-priority queue)*/\\
$~~~~~~~ws$~~~~/*window size*/\\
$~~~~~~~rs$~~~~/*required cache space*/\\
\textbf{local variables:}\\
$~~~~~~~ecd$~~/*empty cache database (recency-keyed min-priority queue)*/\\
$~~~~~~~oc$~~~~/*an object cluster of lru objects\\
$~~~~~~~~~~~~~~~~$(profit-keyed min-priority queue of evictable objects)*/\\
$~~~~~~~co$~~~~/*a candidate object to be included in a cluster*/\\
$~~~~~~~ts \leftarrow 0$~~~/*total size of $ws$ objects initialized to zero*/\\
$~~~~~~~eo$~~~~/*an evicted object*/\\
$~~~~~~~c \leftarrow 0$~~~~/*counter for objects in a cluster initialized to zero*/\\
\textbf{begin}\\
$~~~~~~~$\textbf{if}~~~~$cd.getTotalNumberOfObjects() < ws$\\
$~~~~~~~$\textbf{then} $ws \leftarrow cd.getTotalNumberOfObjects()$;\\
$~~~~~~~ecd \leftarrow cd$;\\
$~~~~~~~$\textbf{do}\\
$~~~~~~~~~~~~~co \leftarrow ecd.removeLeastRecentlyUsedObject()$;\\
$~~~~~~~~~~~~~ts \leftarrow ts + co.getSize()$;\\
$~~~~~~~~~~~~~oc.addObject(co)$;\\
$~~~~~~~~~~~~~c \leftarrow c + 1$;\\
$~~~~~~~\textbf{while~} (c < ws) \vee (ts < rs)$;\\

$~~~~~~~$\textbf{do}\\
$~~~~~~~~~~~~~eo \leftarrow oc.removeMinProfitObject()$;\\
$~~~~~~~~~~~~~cd.evict(eo)$;\\
$~~~~~~~$\textbf{while~} $cd.getFreeSpace() < rs$;\\
\textbf{return}~$cd.getFreeSpace()$;\\
\hline
\end{tabular}
\end{footnotesize}
\end{table}

A significant parameter influencing all performance aspects of i-Cloud is window size ($ws$ in Table\,\ref{i-CloudAlgorithm}).  Our experiments indicated that window size varied at least from one workload to another. Seeking an optimal window size is not an easy task. Lessons learned from our previous studies \cite{cloudcache1,cloudcache2,cloudcache3} indicated that both current cache state, influenced by preceding object requests (i.e., there existed correlations in the request streams) together with allocated cache size, and required cache space were main factors to window size optimization. Therefore, i-Cloud has integrated a multilayer perceptron (MLP) as a forecaster component to automatically learn both of the main factors to forecast near-optimal window sizes to be input into $ws$ in Table\,\ref{i-CloudAlgorithm}. 

\begin{figure}[!tp]
\centering
\includegraphics[width=3.5in]{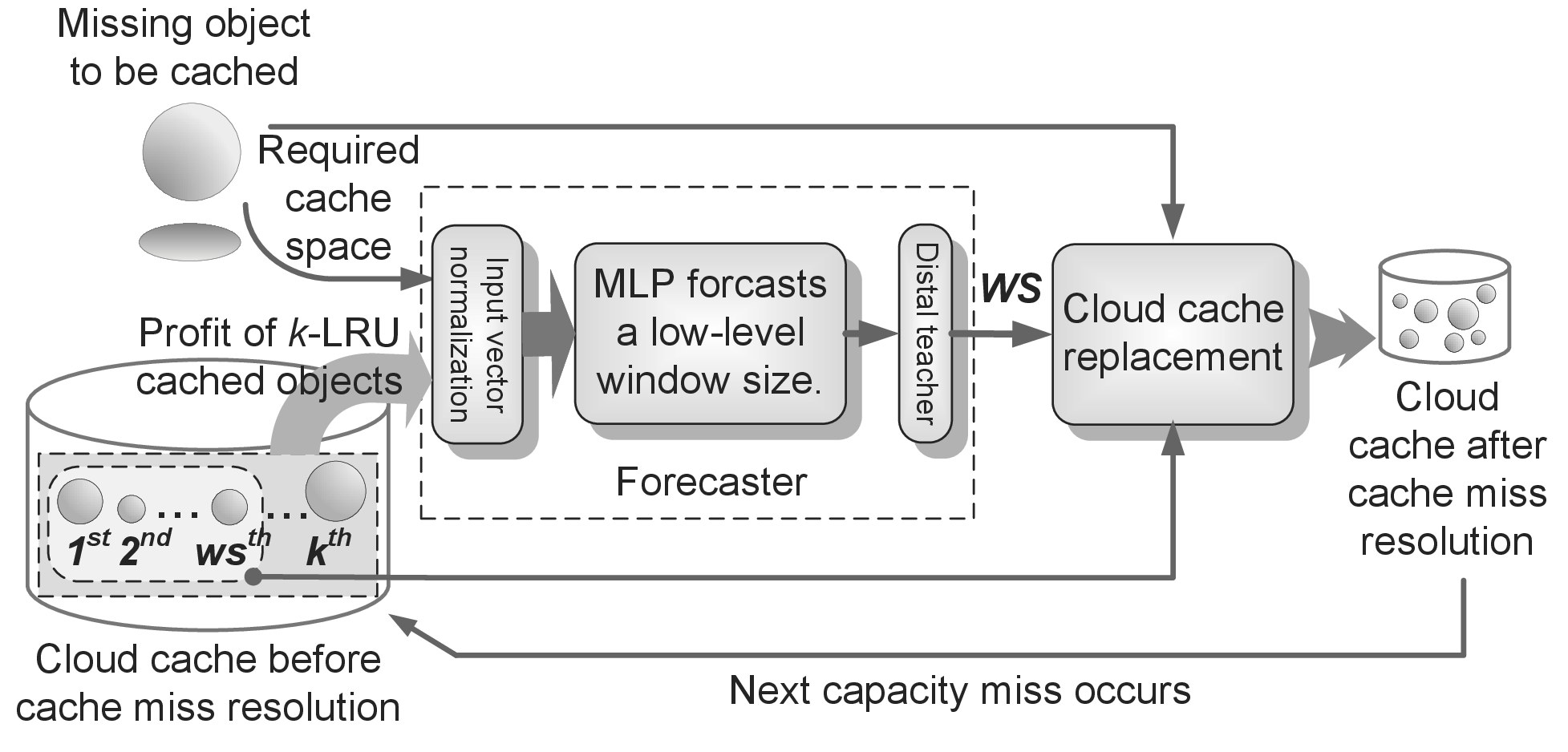}
\caption{Conceptual framework of i-Cloud.}
\label{i-CloudwithMLP}
\end{figure}

Fig.\,\ref{i-CloudwithMLP} demonstrates the conceptual framework of i-Cloud. It consists of two main processing modules, forecaster and cloud cache replacement. They operate as follows. When a capacity miss (i.e., a request for an object that was in a cache but has been since purged, thus cache eviction is required to serve the request) takes place, the vector of $k$ lru cached objects' profits together with a required cache space are fed into the forecaster to forecast a near-optimal window size. During this stage, each input vector is passed internally into the input vector normalization process then the normalized vector is presented to the MLP component to forecast a low-level window size. Such a low-level window size is later denormalized by the distal teacher component to obtain a practical (and potentially near-optimal) window size. The practical window size is subsequently presented to the cloud cache replacement module, which follows i-Cloud algorithm in Table\,\ref{i-CloudAlgorithm}.

Regarding MLP structure that prescribes the number of layers and nodes to be placed in each layer of the MLP, we aimed at the smallest structure possible since too large MLP learns training set well but is unable to generalize. We have designed the MLP structure based on the following guidelines.
\begin{itemize}
\item The appropriate number of input nodes were determined by two lessons learned from our previous studies \cite{cloudcache1}, \cite{cloudcache2}, \cite{cloudcache3}. First, both current cache state, which is influenced by past object request stream together with total allocated cache size, and required cache space were main factors to window size optimizations. This design guideline has been implemented as shown in Fig.\,\ref{i-CloudwithMLP}: a current cache state is captured in the form of a profit vector while a missing object size is used to indicate a minimum required cache space. Second, the optimal window sizes ranged between 100 and 8,000 meaning that the lowest profitable objects to be purged from cache were potentially found inside a cluster of 8,000 lru cached objects. Leveraging this fact simplifies and practicalizes the MLP's implementation since a current cache state can be snapshot by examining only the part of cache database rather than the whole one, which can be much more time consuming. However, to give chance for any unveiled maximum values of optimal window sizes, we simply chose 10,000 (i.e., $k$ in Fig.\,\ref{i-CloudwithMLP}) lru cached objects to represent each current cache state to be examined by the MLP. Thus, the total number of input nodes is 10,002 including a required cache space node and a bias one.
\item It is conventionally guided that some continuous functions cannot be approximated accurately by single-hidden-layer MLP whereas two hidden layers are sufficient to approximate any desired bounded continuous function, which is also the case of i-Cloud. Moreover, the hidden layers are usually kept at approximately the same size to ease training. 
\item The output layer necessitates a single node to deliver an estimated low-level window size. 
\end{itemize}

Several structures were experimented in an effort based on these guidelines. We finally came out at a minimal structure that can be expressed in the conventional notation of 10,002/2/2/1, which means 10,002 input nodes, two hidden layers with two nodes each, and a single output node. All the nodes are fully-connected between two adjacent layers, except a bias node is connected to all noninput ones. A complete input vector presented to the MLP is denoted as 
\begin{equation}
\label{vectorX}
X = <1, p_1, p_2, ..., p_{10,000}, rs>
\end{equation}
\setlength{\parindent}{0in}where $p_{i=1~to~10,000}$ is the profit of $i^{th}$ lru object and $rs$ is a required~cache~space. Notice that the bias is set to a constant activation 1. Every link connecting node $j$ to node $i$ has an associated weight $w_{ij}$.\parindent    1.0em

\subsection{Learning phase}
\label{Learning phase}
Besides the MLP structure, to obtain the complete MLP mandates the appropriate set of weights on all node-connecting links. This has been done by means of supervised learning with a distal teacher and back-propagation. We used mean squared error as an objective function. Our MLP learns patterns inherent in caching state history. Each input pattern is organized into a training vector of the form $X$ (Eq.(\ref{vectorX})) and was generated every time capacity miss occurs during the i-Cloud simulation of a certain trace, cache size and respective (byte-hit or uniform-cost-saving) optimal window size. The result of each complete simulation session was a sequence of input patterns, which were contained inside a single training data set. We generated two distinct data sets by the separate simulation sessions of two 15-day traces (Table\,\ref{traces}) using 10\% cache size (i.e., percent of the maximum bytes of total unique objects referenced in each trace). The first data set, BO15D10\%, contained totally 54,639 patterns, while the other data set, NY15D10\%, contained 62,607 patterns. We normalized every element value within each input vector except the bias constant by max-min linear scaling. Because the MLP outputted a low-level window size of domain [0.0, 1.0] via sigmoid function, to achieve desired window sizes requires a distal teacher to denormalize the low-level window size to obtain an actual window size of [0, 10,000]. Target window sizes were the static optimal ones aforementioned. Note that any actual window size near the lower bound tended to be increased later by i-Cloud as coded in Table\,\ref{i-CloudAlgorithm} to be able to fit a required cache space. We used adaptive learning rate instead of fixed one to make the forecaster converge quickly. The learning phase terminated when mean squared error stabilized below 1000.

\subsection{Validation phase} 
\label{Validation phase}
Our previous work \cite{cloudcache4} has shown that measuring the accuracy of the trained forecaster by using validation data sets to measure errors produced by the forecaster was almost meaningless as a local optimum window size could have the high degree of validation error. Instead, we have evaluated the performances of i-Cloud including the trained forecaster as a whole by trace-driven simulations in the four performance metrics as described in Section \ref{Results and discussions}.

\begin{table}[!t]
\renewcommand{\arraystretch}{1.25}
\caption{Forecaster algorithm}
\label{ForecasterAlgorithm}
\begin{footnotesize}
\begin{tabular}{l}
\hline
\textbf{algorithm:} \texttt{Forecaster}\\
\hline
\textbf{input variables:}\\
$~~~~~~~~cd$~~~~~~~~~/*cache database (recency-keyed min-priority queue)*/;\\
$~~~~~~~~rs$~~~~~~~~~/*required cache space*/;\\
$~~~~~~~~w_{ij}$~~~~~~~/*inter-node connection weight*/;\\
\textbf{local variables:}\\
$~~~~~~~~ecd$~~~~~~~/*empty cache database*/;\\
$~~~~~~~~ip$~~~~~~~~~/*MLP's input node*/;\\
$~~~~~~~~i,~j$~~~~~~~/*node indices where j is on preceding layer of i*/;\\
$~~~~~~~~temp$~~~~/*temporary object*/;\\
$~~~~~~~~l$~~~~~~~~~~~/*layer index*/;\\
$~~~~~~~~u_i$~~~~~~~~~/*weighted sum*/;\\
$~~~~~~~~y_i$~~~~~~~~~/*output activation*/;\\
\textbf{begin}\\
$~~~~~~~~ecd \stackrel{1\ldots10000}{\longleftarrow} cd$;~~/*~duplicating 10,000 lru cached objects~*/\\
$~~~~~~~~$/*~Read a current cache state into 10,002 input nodes~*/\\
$~~~~~~~~ip_0 \leftarrow 1$; /*~bias constant~*/\\
$~~~~~~~~$\textbf{for} $j$=1 to 10000~\textbf{do}\\
$~~~~~~~~~~~~~~temp \leftarrow ecd.removeLeastRecentlyUsedObject()$;\\
$~~~~~~~~~~~~~~ip_j \leftarrow temp.getProfit()$;\\
$~~~~~~~~ip_{10001} \leftarrow rs$;\\
$~~~~~~~~$/*~Normalize inputs except bias node~*/\\
$~~~~~~~~$\textbf{for} $j$=1 to 10001~\textbf{do}\\
$~~~~~~~~~~~~~~ip_j \leftarrow 2~ip_j~/~10^{14}~-~1$;\\
$~~~~~~~~$/*~Propagate the inputs forward to compute an output~*/\\
$~~~~~~~~$\textbf{for each} $noninput~layer~\ell$ \textbf{do}\\
$~~~~~~~~~~~~~~$\textbf{for each} $node~i~in~\ell$ \textbf{do}\\
$~~~~~~~~~~~~~~~~~~~~u_i \leftarrow \sum_{j}w_{ij}ip_j$;\\
$~~~~~~~~~~~~~~~~~~~~y_i \leftarrow sigmoid(u_i)$;\\
$~~~~~~~~$/*~Denormalize the output~*/\\
$~~~~~~~~y_{10006} \leftarrow 10000~y_{10006}$;\\
$\textbf{return}~y_{10006}$;~~/*return window size*/\\
\hline
\end{tabular}
\end{footnotesize}
\end{table}

Table\,\ref{ForecasterAlgorithm} presents the forecaster's pseudo code, basically comprising an input vector normalization, a feedforward propagation and an output denormalization (i.e., a distal teacher) parts, respectively. To become an effective forecaster, it is mandatory to train the MLP as described in the following subsection. 

\begin{figure*}[!t]
\centering
\includegraphics[width=7.2in]{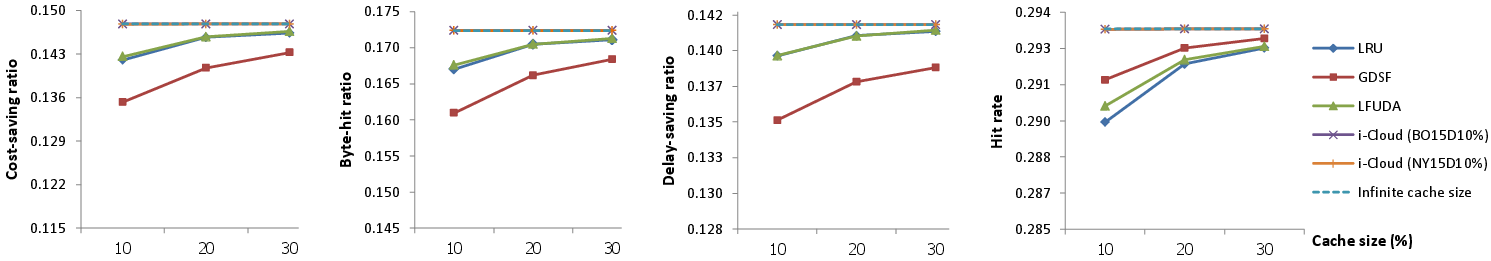}
\caption{The comparative performance results of i-Cloud and various approaches based on the 31-day BO trace.}
\label{31dBOperformances}
\end{figure*}

\begin{figure*}[!t]
\centering
\includegraphics[width=7.2in]{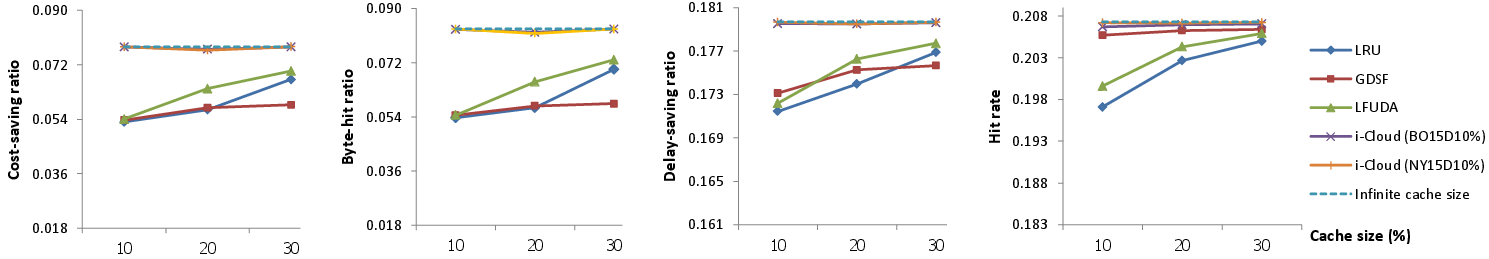}
\caption{The comparative performance results of i-Cloud and various approaches based on the 31-day NY trace.}
\label{30dNYperformances}
\end{figure*}

\subsection{Algorithmic practicality}
\label{Algorithmic practicality}
Because the total number of cached objects (N) can be increasingly large at runtime. It is necessary to conduct the algorithmic practicality analysis of both modules in Fig.\,\ref{i-CloudwithMLP}. As for the time complexity analysis of i-Cloud algorithm (Table\,\ref{i-CloudAlgorithm}), the statements that take significant part in processing time are: replicating $cd$ into $ecd$ takes O(NlogN); the first $do-while$ loop takes O(NlogN) as the window size can be set via the preceding $if-then$ statement to as many as N and removing each object from $ecd$ is O(logN) just like adding each object into $oc$; the second $do-while$ loop has the worst-case running time of O(NlogN) because the number of evicted objects is bounded by N, while removing each object from $oc$ takes O(logN) and deleting an object from $cd$ is O(logN). The other statements are all identically O(1). Therefore, the algorithm is O(NlogN).

Regarding the time complexity of the forecaster algorithm (Table\,\ref{ForecasterAlgorithm}): copying 10,000 lru objects from $cd$ to $ecd$ takes O(logN), the first $for$ loop takes O(logN) to remove lru object from $cd$, while the other statements are all identically O(1). Thus, the algorithm is O(logN).

Since the cloud cache replacement and forecaster modules are sequentially connected as shown in Fig.\,\ref{i-CloudwithMLP}, i-Cloud algorithm is totally O(NlogN) + O(logN) equal to O(NlogN). In other words, i-Cloud strategy can be implemented. 

\subsection{Monetary cost models}
\label{Monetary cost models}
Since this paper aims for multi-provider hybrid clouds, we engaged a nonuniform cost model for evaluation. In nonuniform cost model, all of the 50 domains (Cf. Section \ref{Original trace data sets}) of each trace were assumed to go to a hybrid cloud, running in two distinct public clouds that offered different data-out charge rates. The first charge rate was set to 0.1535 USD/GB based on Google Cloud Storage (Cf. Section \ref{Introduction}), while the other was set to 0.0829 USD/GB based on Amazon S3's data transfer out to Internet charge (US Standard) as of September 2013 \cite{AWS}. (Both flat rates are actually the weighted averages of the actual regressive rates of Google Cloud Storage and Amazon Simple Storage Service covering the monthly data transfer of the representative scenario described in Section \ref{Introduction}.) Both flat rates were associated with the 50 domains of each trace in an interleaving manner so that the nonuniform cost model could be emulated for each trace. Objects retrieved from the same domain were always charged at the same rate. 

As a remark, although the evaluation relied on the nonuniform cost model, it was our intention to train i-Cloud by using static optimal window sizes, which were tuned based on a uniform cost model rather than nonuniform one. This was to observe how efficiently i-Cloud learning uniform-cost data sets performed against nonuniform-cost ones.

\section{Results and discussions}
\label{Results and discussions}
The one-month traces in Table\,\ref{traces} were used to conduct the separate trace-driven simulations of i-Cloud learning BO15D10 data set, i-Cloud learning NY15D10 data set and the other three well-known approaches, LRU, GDSF, and LFU-DA, supported by popular Squid caching proxy \cite{Squid}. The comparative simulation results in four performance metrics are demonstrated in Fig.\,\ref{31dBOperformances} and \,\ref{30dNYperformances}. The simulated cache sizes are dictated in percents of the maximum bytes of total unique objects of each of the one-month traces. Interesting observations are as follows.

\begin{itemize}
\item It is obvious that i-Cloud of both learning data sets has outperformed all competitive approaches in all simulation cases in not only cost-saving but also byte-hit, delay-saving and hit performance metrics. This means that though i-Cloud learned uniform cost patterns, it has delivered good performances in nonuniform cost environment.
\item i-Cloud performances have shown to stabilize at close degrees to those of infinite cache size in nonuniform cost environment. This is also the strength of i-Cloud against uniform cost models that has been known in our previous works \cite{cloudcache1,cloudcache2,cloudcache3,cloudcache4}. 
\item i-Cloud that learned only small data set derived from 15-day trace has still been able to perform relatively well against the one month workloads of both user communities. This substantiates the successful long-term deployment of i-Cloud up to some extent. 
\item Based on the BO trace and cost-saving metric, i-Cloud(BO15D10\%) has outperformed i-Cloud(NY15D10\%) at 10\% cache size and delivered identical performance to i-Cloud(NY15D10\%) at 20\% and 30\% cache sizes. Using the NY trace, i-Cloud(NY15D10\%) has more economized than i-Cloud (BO15D10\%) at 10\% and 30\% cache sizes and less economized than i-Cloud (BO15D10\%) at 20\% cache size. Hence, in most cases i-Cloud has performed economically better against user community behavior i-Cloud has learned.
\item i-Cloud could retain both high byte-hit and hit rates at the same time based on our simulated workloads. This means that the breakthrough finding of our previous works (\cite{cloudcache2, cloudcache3, cloudcache4}) that optimal hit and byte-hit ratios could be attained at the same time also applies to nonuniform cost environment. Since i-Cloud tends to evict smaller objects, the finding is an objection to the conventional rule of thumb in traditional web caching: ``strategies that tend to remove bigger objects improve the hit-rate but decrease the byte-hit-rate" \cite{Podlipnig}.
\end{itemize}

\section{Conclusion}
This paper presents i-Cloud cache eviction approach that accommodates the distributed sharings of big data. i-Cloud has access recency as a priority factor for object replacement decision. i-Cloud parameterizes an MLP-based self-tuning window size to generalize the recencies of objects within a formulated object cluster. The lowest profitable clustered objects are purged from cloud cache. Based on the trace-driven simulation results, the distributed sharings of big data was most efficient when employing i-Cloud. Although i-Cloud has been trained based on a uniform cost model, it performed well against a nonuniform cost environment or multi-provider hybrid cloud.

\section*{Acknowledgment}
This research is financially supported by Thailand's Office of the Higher Education Commission, Thailand Research Fund, and Sripatum university (grant MRG5580114). The authors also thank Duane Wessels, National Science Foundation (grants NCR-9616602 and NCR-9521745) and the National Laboratory for Applied Network Research for the trace data used in this study.

\bibliographystyle{IEEEtran}
\bibliography{CLOUD}
\end{document}